\documentclass[english,aps,prb,twocolumn,amsmath,amssymb,showpacs,superscriptaddress,notitlepage,longbibliography]{revtex4-2}

\usepackage[T1]{fontenc}
\usepackage{color}
\definecolor{page_backgroundcolor}{rgb}{1, 1, 1}
\pagecolor{page_backgroundcolor}
\usepackage{amsmath}
\usepackage{graphicx}
\usepackage{microtype}
\usepackage{bm}
\usepackage{makecell}
\usepackage[normalem]{ulem}
\usepackage{enumitem}
\usepackage[marginal]{footmisc}
\usepackage{diagbox}

\makeatletter
\usepackage{amssymb}
\usepackage{graphicx}
\usepackage[colorlinks=true,linkcolor=blue,anchorcolor=red,citecolor=blue,urlcolor=blue]{hyperref}
\usepackage[caption=false]{subfig}
\usepackage{lipsum}

\makeatother

\usepackage{babel}
\renewcommand{\appendixname}{\MakeUppercase{\appendixname}}
\begin{document}
\global\long\def\figurename{Fig.}

\title{Field-free Josephson diode and tunable $\phi_0$-junction in chiral kagome antiferromagnets}
\author{Jin-Xing Hou }
\thanks{These authors contributed equally.}
\affiliation{Hefei National Laboratory, Hefei, 230088, China}

\author{Chuang Li}
\thanks{These authors contributed equally.}
\affiliation{Center for Correlated Matter and School of Physics, Zhejiang University, Hangzhou 310058, China}
\affiliation{Hefei National Laboratory, Hefei, 230088, China}

\author{Lun-Hui Hu}
\email{lunhui@zju.edu.cn}
\affiliation{Center for Correlated Matter and School of Physics, Zhejiang University, Hangzhou 310058, China}

\author{Song-Bo Zhang}
\email{songbozhang@ustc.edu.cn}
\affiliation{Hefei National Laboratory, Hefei, 230088, China}
\affiliation{School of Emerging Technology, University of Science and Technology of China, Hefei, 230026, China}

\date{\today}

\begin{abstract}

The recent realization of superconducting proximity effect in chiral antiferromagnets (cAFMs) opens a new route to nonreciprocal superconducting transport of fundamental interest and practical importance. Using microscopic modeling and symmetry analysis, we show that Josephson junctions formed by conventional $s$-wave superconductors (SCs) and cAFMs on the kagome lattice exhibit Josephson diode effects and anomalous phase shifts ($\phi_0$-junction state)  when space inversion $\mathcal{I}$, time-reversal $\mathcal{T}$, and combined mirror-time-reversal $\mathcal{TM}_z$ symmetries are simultaneously broken. We propose two setups to realize these phenomena and achieve high diode efficiency. (i) An SC/cAFM/SC junction with spin–orbit coupling, which enables a field-free diode effect with a robust tunable $\phi_0$-junction state. (ii) An SC/cAFM/cAFM$^\prime$/SC junction, where two cAFM layers with different in-plane order orientations, under an out-of-plane Zeeman exchange field, produces significant diode effect and anomalous phase shifts. 
These results establish a direct link between $\mathcal{TM}_z$ symmetry breaking and nonreciprocal superconductivity, suggesting cAFMs as versatile platforms for symmetry-engineered Josephson diodes and tunable $\phi_0$-junctions.

\end{abstract}

\maketitle


\newpage
\noindent

\newpage

\section{Introduction}

The dc Josephson effect is a macroscopic quantum phenomenon in which supercurrents flow between two superconductors (SCs) without an applied voltage~\cite{Tinkham2004Book}. It stems from phase-coherent tunneling of Cooper pairs and underlies a broad range of superconducting phenomena and technologies. In conventional junctions that preserve time-reversal and inversion symmetries, the current-phase relation is odd, $I(\phi)=-I(-\phi)$, which enforces equal forward and backward critical currents, $I_c^\rightarrow=I_c^\leftarrow$. When these symmetries (e.g., time-reversal and inversion) are broken, the current-phase relation can become asymmetric under $\phi\rightarrow -\phi$ and a Josephson diode effect may emerge, $I_c^\rightarrow\neq I_c^\leftarrow$~\cite{hu2007PRL,davydova2022universal,zhang2022general,nadeem2023superconducting,he2022phenomenological,misaki2021theory,tanaka2022theory,jiang2022superconducting,wang2025current,wang2025josephson,wang2025universal}. 
This nonreciprocal superconducting transport enables dissipationless rectification~\cite{amundsen2024colloquium,debnath2024gate,moll2023evolution} and is driving growing interest for {\color{black}quantum information processing~\cite{golod2022demonstration,gupta2023gate,souto2022josephson,nikodem2025tunable,liu2024josephson,ando2020observation} and superconducting spintronic devices}~\cite{he2023supercurrent,yuan2022supercurrent,DaidoPRL2022,nagaosa2024nonreciprocal,HouPRL2023,lyu2021superconducting,ma2025superconducting,narita2022field,liu2024superconducting,costa2025unconventional,schulz2025theory,schulz2025quantum,nikolic2025spin,djurdjevic2025josephson,PRB2024Superconducting,reinhardt2024link}.

In Josephson junctions, a closely related phenomenon is the $\phi_0$-junction~\cite{guo2025phi,lu2024varphi,buzdin2003periodic}, where the ground-state phase difference $\phi_0$ deviates from $0$ or $\pi$ as a consequence of simultaneous breaking of inversion and time-reversal symmetries. Such phase shifts provide sensitive probes of underlying symmetries and offer opportunities for superconducting qubit~\cite{tafuri2019fundamentals,guo2022qubit,amundsen2024colloquium}, phase batteries~\cite{amundsen2024colloquium,GolodPRL2010,SickingerPRL2012,assouline2019spin,goldobin2013memory,strambini2020josephson}, \emph{etc}. Josephson diode and $\phi_0$-junction state have been explored in diverse platforms, including Rashba spin-orbit coupled systems~\cite{baumgartner2022effect,debnath2024gate,bhowmik2025optimizing,MondalPRB2025}, topological materials~\cite{liu2024josephson,pal2022josephson,lu2023tunable,tanaka2022theory,fracassi2025intrinsic,Chen2018PRB}, van der Waals heterostructures~\cite{wu2022field,qiu2023emergent,hu2025polarity,Hu2023PRL}, as well as the recently discovered collinear altermagnets~\cite{chakraborty2025perfect,cheng2024field,BanerjeePRB2024,Sharma2025tunable,Boruah2025PRB,debnath2025spin,sahoo2025field}. Despite extensive theoretical~\cite{hu2007PRL,davydova2022universal,zhang2022general,jiang2022superconducting,souto2022josephson} and experimental progress~\cite{HouPRL2023,ando2020observation,Sickinger2012experimental}, a unified symmetry-based framework for tunable Josephson diodes and $\phi_0$-junctions within microscopic models remains elusive. 

Recently, the superconducting proximity effect has been realized in chiral antiferromagnets (cAFMs) on kagome lattice, most notably in Mn$_3$Ge thin films~\cite{jeon2021long,jeon2023chiral}. 
These materials host robust noncollinear order in the kagome plane while exhibiting zero net magnetization, leading to significant nonrelativistic spin splitting in the electronic bands~\cite{Chen2014PRL,kiyohara2016giant,nakatsuji2015large,zhang2017strong,ikhlas2017large,LHHu25SCPMA,SBZ25PRB}. cAFMs are broadly regarded as a distinct, non-collinear type of altermagnets~\cite{cheong2025altermagnetism}. They display exotic quantum transport properties such as giant anomalous Hall effect~\cite{Chen2014PRL,kiyohara2016giant,nakatsuji2015large,zhang2017strong}, anomalous Nernst effect~\cite{cheng2016spin,narita2017anomalous,ikhlas2017large}, and current-induced spin torques~\cite{shao2021roadmap,baltz2018antiferromagnetic,kimata2019magnetic,holanda2020magnetic,liu2019current}.  
Moreover, their peculiar spin texture and spin splitting have been shown to promote pronounced spin-triplet pairing that contributes substantially to the supercurrent~\cite{SBZ25newton,hou2025enhancement,chaou2025proximity}. While the bulk cAFMs are inversion symmetric, practical cAFM-based hybrid junctions naturally break inversion at interfaces due to kagome lattice geometry and structural asymmetry. 
Despite these favorable ingredients and experimental feasibility, their potential for nonreciprocal Josephson transport has yet to be systematically investigated.

To this end, we study Josephson junctions composed of conventional $s$-wave SCs and cAFMs on kagome lattice. We find that breaking the combined mirror–time-reversal symmetry $\mathcal{TM}_z$ (with $\mathcal{T}$ time-reversal and $\mathcal{M}_z$ the mirror about the kagome plane) is a key requirement for the Josephson diode effect beyond the breaking of time-reversal and inversion alone. 
This symmetry principle provides a general guideline for engineering nonreciprocal Josephson responses in kagome-based cAFM systems. To illustrate this, we propose two experimentally feasible device implementations, as sketched in Fig.~\ref{fig:setup}. The first setup is an SC/cAFM/SC junction in which spin–orbit coupling (SOC) is introduced to explicitly break $\mathcal{TM}_z$ and inversion $\mathcal{I}$ at the junction interfaces. This enables a field-free Josephson diode and a tunable $\phi_0$-junction. The second setup is an SC/cAFM/cAFM$'$/SC junction operated with an out-of-plane Zeeman exchange field (or spin canting), where the relative orientation of the cAFM orders in the two cAFM layers controls sizable diode responses and anomalous phase shifts. Notably, these effects do not occur in ferromagnetic counterparts, indicating the essential role of cAFM order. 
Our results suggest that  
combining kagome geometry with chiral magnetic textures and controlled symmetry breaking provides a promising route toward realizing Josephson diodes and related superconducting spintronic functionalities.


\section{Results}

\subsection{Model Hamiltonian and symmetry analysis}\label{sec:setup_model}

\begin{figure}[t]
\centering
\includegraphics[width=1\linewidth]{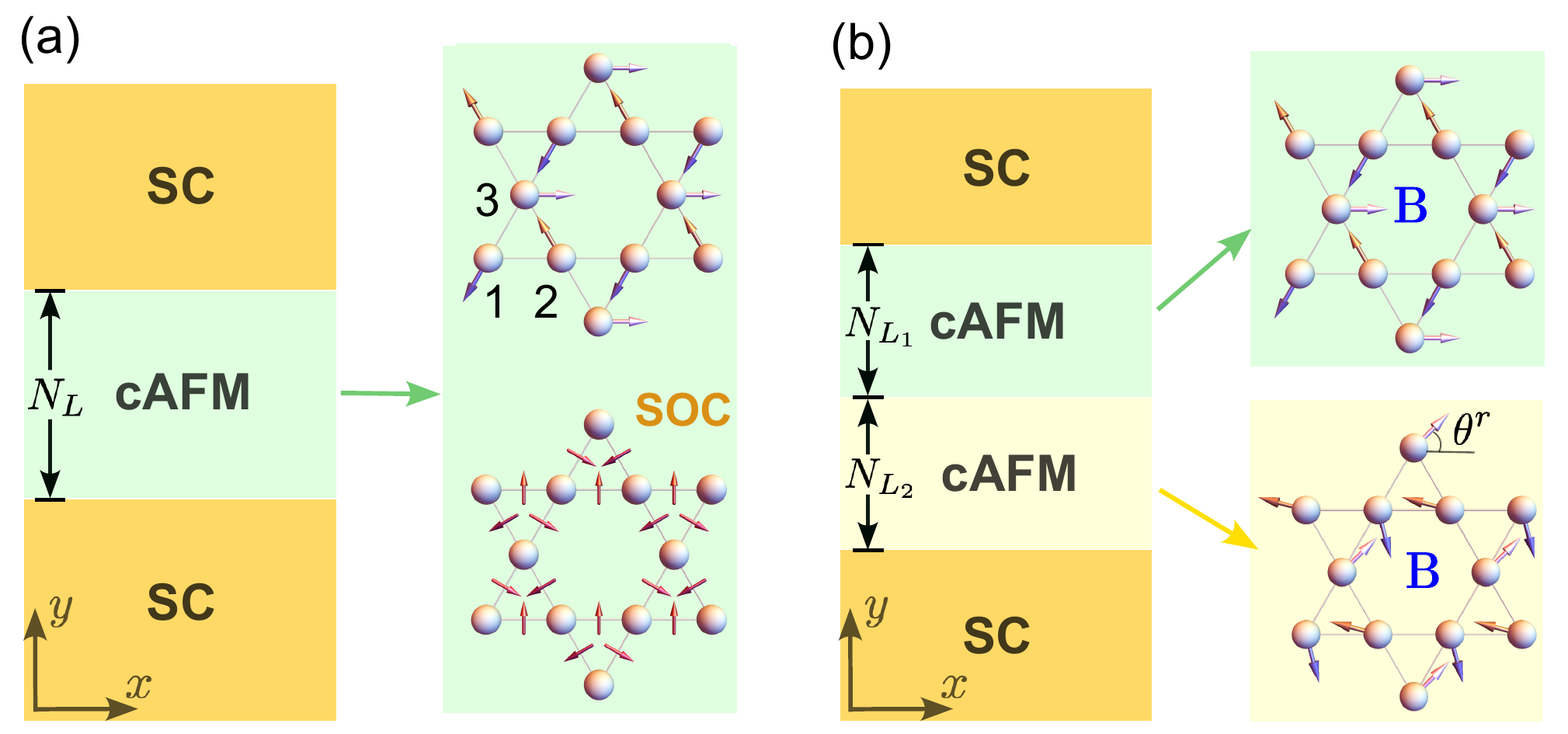}
\caption{Schematic of two Josephson setups based on cAFM on the kagome lattice. (a) The SC/cAFM/SC junction with SOC: the site arrows show the local magnetic moments, and the bond arrows mark the coplanar unit vectors $\bm{n}_{\mu\nu}$ associated with SOC in the cAFM. (b) The SC/cAFM/cAFM$'$/SC junction with two cAFM layers in middle: the two noncollinear orders differ by a relative angle $\theta^r$ and an out-of-plane Zeeman field $B_z$ is applied there.
}
\label{fig:setup}
\end{figure}

We start with the SC/cAFM/SC Josephson junction, in which a cAFM is sandwiched between two conventional $s$-wave superconducting leads [see Fig.~\ref{fig:setup}(a)]. The junction is oriented along the $y$-direction and translational symmetry is assumed in the $x$-direction. Thus, the momentum $k_x$ remains a good quantum number and the system can be treated as a stack of “rows” of triangular magnetic unit cells along the $y$-direction. 
The total Hamiltonian of the junction is given by  
\begin{equation}   
H_{\text{SNS}}=H_0+H_\Delta+H_{\text{cAFM}}+H_{so},
\label{eq:total-H}
\end{equation}
where $H_0$ is the prototypical tight-binding Hamiltonian on the kagome lattice 
\begin{equation} 
    H_0=  \sum_{k_x,l} \big\{\Psi_{\ell}^\dagger \mathcal{H}(k_x) \Psi_{\ell}^{\vphantom{\dagger}} + \big[\Psi_{\ell}^\dagger \mathcal{V}(k_x) \Psi_{\ell+1}^{\vphantom{\dagger}}+\mathrm{h.c.}\big]\big\},
    \label{eq:H0}
\end{equation}
with the spinor basis $\Psi_{\ell}^\dagger = (c_{\ell,1,\uparrow}^\dagger, c_{\ell,2,\uparrow}^\dagger, c_{\ell,3,\uparrow}^\dagger,c_{\ell,1,\downarrow}^\dagger, c_{\ell,2,\downarrow}^\dagger$, $ c_{\ell,3,\downarrow}^\dagger)$.  The operator
$c_{\ell,\nu,\sigma}^\dagger$ creates an electron with spin $\sigma\in\{\uparrow,\downarrow\}$ at sublattice $\nu \in \{1,2,3\}$ of the $\ell$-th row of unit cells along the $y$-direction. 
Here, for ease of notation, we suppress $k_x$ dependence in the basis $c_{\ell,\nu,\sigma} \equiv c_{k_x, \ell,\nu,\sigma}$. $\text{h.c.}$ indicates the Hermitian conjugate of the preceding term.
$\mathcal{H}(k_x) = s^0 h(k_x)$ and $\mathcal{V}(k_x)= s^0 V(k_x)$ correspond to the local term for each row and the hoping matrix between neighboring rows, respectively. 
$s^0$ is the unit matrix and $\bm{s} = (s^x,s^y,s^z)$ is the Pauli matrix vector in spin space. The matrices $h(k_x)$ and $V(k_x)$ are 
\begin{subequations}
\begin{align}
    h(k_x) = & -t  \left(\begin{matrix}
        {\mu}/{t}-1 & 1+ e^{-ik_x} & 1 \\
        1+ e^{ik_x} & {\mu}/{t}-1 & 1 \\
        1 & 1 & {\mu}/{t}-1 \\
    \end{matrix}\right), \\
    V(k_x) = & - t \left(\begin{matrix}
        0 & 0 & 0 \\
        0 & 0 & 0 \\
        1 & e^{-ik_x} & 0 \\
    \end{matrix}\right),
    \label{eq:Tmatrix}
\end{align}
\end{subequations}
where $t$ is the hopping amplitude between neighboring lattice sites, and $\mu$ is the chemical potential. We set the lattice constant $a=1$, $t$ to be the unit of energy, and the Dirac points of the kagome model at zero energy.  

The second term $H_{\Delta}$ in Eq.~\eqref{eq:total-H} describes the $s$-wave pairing potential in the superconducting leads
\begin{equation}
 H_{\Delta} = \sum_{\ell}  \Delta_\ell [\Psi_{\ell,\uparrow}^\dagger (\Psi_{\ell,\downarrow}^\dagger)^T -\Psi_{\ell,\downarrow}^\dagger (\Psi_{\ell,\uparrow}^\dagger)^T]+ \mathrm{h.c.},
 \label{eq:pairing}
\end{equation}
where $\Psi_{\ell,\sigma}^\dagger=(c_{\ell,1,\sigma}^\dagger,  c_{\ell,2,\sigma}^\dagger, c_{\ell,3,\sigma}^\dagger)$ with $\sigma\in\{\uparrow,\downarrow\}$, and  $\Delta_\ell=\Theta(\ell+N_L/2)\Delta_0 e^{-i\phi/2} + \Theta(\ell-N_L/2)\Delta_0 e^{i\phi/2}$. $\Delta_0$ is the magnitude of the pairing potential, $\phi$ is the phase difference between the two SCs, 
$\Theta(x)$ is the Heaviside step function, and $N_L$ is the distance between two SCs (i.e., the length of cAFM region) in units of row spacing $\sqrt{3}a$. 
The third term in Eq.~\eqref{eq:total-H} is an on-site exchange describing the non-collinear magnetic order in the junction
\begin{align}\label{eq:H_cAFM}
 H_{\text{cAFM}} &= \sum_{\ell,\nu}  \Psi^\dagger_{\ell,\nu}\ \bm{m}_{\ell,\nu}^{\vphantom{\dagger}} \cdot \bm{s}\ \Psi_{\ell,\nu}^{\vphantom{\dagger}},  
\end{align}
where $\bm{m}_{\ell,\nu} = J_\ell (\cos{\theta_\nu},\sin{\theta_\nu},0)$ denotes the local magnetic moment on the $\nu$-sublattice ($\nu\in\{1,2,3\}$) in the $\ell$-th row of unit cells. The exchange strength is given by $J_\ell=J\Theta(\ell+N_L/2+1)\Theta(-\ell+N_L/2+1)$ and in-plane angles $\theta_\nu$ specify the in-plane magnetic configuration. The sublattice moments $\bm{m}_\nu$ form 120$^\circ$ in-plane pattern [see Fig.~\ref{fig:setup}], which we parameterize them as $\theta_{2/3}=\theta_{1}\pm 2\pi/3$ or $\theta_{2/3}=\theta_{1}\mp 2\pi/3$. 

In the kagome antiferromagnet, SOC may exist intrinsically, as evidenced by large anomalous Hall effects in the materials~\cite{nakatsuji2015large,yin2022topological,yin2018giant,liu2025multipolar,liu2022PRXspingroup}. 
It can be described by~\cite{Chen2014PRL}
\begin{align}
    H_{so} = i t_{so} \sum_{\langle \cdots\rangle, \sigma,\sigma^\prime}\lambda_{\mu\nu} \bm{n}_{\mu\nu} \cdot  \bm{s}_{\sigma\sigma^\prime} c_{j,\ell,\mu,\sigma}^\dagger c_{j^\prime,\ell^\prime,\nu, \sigma^\prime}^{\vphantom{\dagger}}, 
\end{align}
where $c_{j,\ell,\nu,\sigma}^\dagger$ ($c_{j,\ell,\nu,\sigma}$) creates (annihilates) an electrons on the sublattice $\nu \in\{1,2,3\}$ of the unit cell located at column $j$ and row $\ell$, and the spin indices are $\sigma, \sigma^\prime \in\{\uparrow,\downarrow\}$. 
The notation $\langle\cdots\rangle\equiv\langle j,\ell,\mu; j^\prime,\ell^\prime,\nu\rangle$ restricts the summation to nearest neighbors, $t_{so}$ denotes the SOC strength, and $\lambda_{\mu\nu}=-\lambda_{\nu\mu}$ with $\lambda_{12}=\lambda_{23}=\lambda_{31}=1$. $\bm{n}_{\mu\nu} = \bm{n}_{\nu\mu} = \left(\cos{\varphi_{\mu\nu}},\sin{\varphi_{\mu\nu}},0\right)$ are three coplanar unit vectors on the bond connecting sublattices $\mu$ and $\nu$,
where $\varphi_{12}=\pi/2,\;\varphi_{23}=7\pi/6$ and $\varphi_{31}=-\pi/6$.
In the basis introduced in Eq.~\eqref{eq:H0}, the SOC terms is rewritten as
\begin{align}
H_{so} &= \overline{\sum} \Big[ \sum_{\ell=-\mathcal{L}}^{\mathcal{L}} f_{\mu\nu}(k_x) \; \bm{n}_{\mu\nu} \cdot \bm{s}_{\sigma\sigma^\prime} c_{\ell,\mu,\sigma}^\dagger c_{\ell,\nu, \sigma^\prime}^{\vphantom{\dagger}} \notag \\
&\quad +  \sum_{\ell=-\mathcal{L}}^{\mathcal{L}-1} g_{\mu\nu}(k_x) \; \bm{n}_{\mu\nu} \cdot \bm{s}_{\sigma\sigma^\prime}  c_{\ell,\mu,\sigma}^\dagger c_{\ell+1,\nu, \sigma^\prime}^{\vphantom{\dagger}} \notag \\
&\quad + \sum_{\ell=-\mathcal{L}}^{\mathcal{L}-1} g_{\mu\nu}^\dagger(k_x) \; \bm{n}_{\mu\nu} \cdot \bm{s}_{\sigma\sigma^\prime} c_{\ell+1,\mu,\sigma}^\dagger c_{\ell,\nu, \sigma^\prime}^{\vphantom{\dagger}}\Big], \label{eq:SOC}
\end{align}
where $\overline{\sum}\equiv \sum_{k_x,\mu,\nu, \sigma,\sigma^\prime}$, $\mathcal{L}=\left\lceil N_L/2 \right\rceil$ denotes rounding a number up to the nearest integer of $N_L/2$, the matrix functions $f(k_x)$ and $g(k_x)$ are given by
\begin{subequations}
\begin{align}
f(k_x) = &\; it_{so} \left(\begin{matrix}
        0 & -1-e^{-ik_x}  & -1 \\
        1+e^{ik_x} & 0 & -1 \\
        1 & 1 & 0 \\
    \end{matrix}\right), \\
g(k_x) = &\; i t_{so} \left(\begin{matrix}
        0 & 0  & 0 \\
        0 & 0 & 0 \\
        1 & e^{-ik_x} & 0 \\
    \end{matrix}\right).
\end{align}\label{eq:fg_SOC}
\end{subequations}

In the presence of any symmetry that reverses both the Josephson current direction, $I_s \rightarrow -I_s$, and the superconducting phase difference, $\phi \rightarrow -\phi$, the current-phase relation must satisfy $I_s(\phi) = -I_s(-\phi)$. Thus, to realize a Josephson diode effect, all such symmetries must be broken. In the kagome-cAFM-based Josephson junction, the relevant symmetries include  inversion $\mathcal{I}$, time-reversal $\mathcal{T}$, and the combined mirror-time-reversal symmetry $\mathcal{TM}_z$. 

We first examine inversion symmetry and it microscopic action on the kagome lattice. For concreteness and without loss of
generality, we choose the inversion center at the sublattice `3' of the central row $\ell=0$ in the cAFM. With this choice, inversion acts on the annihilation operators as
\begin{subequations}
\begin{align}
   c_{j,\ell,1,\sigma} \;&\rightarrow c_{-j,-\ell+1,1,\sigma},\;  \\  
   c_{j,\ell,2,\sigma} \;&\rightarrow c_{-j-1,-\ell+1,2,\sigma},\;  \\  
   c_{j,\ell,3,\sigma} \;&\rightarrow c_{-j,-\ell,3,\sigma}.   
\end{align}
\end{subequations}
This transformation leaves the sublattice index unchanged but shifts the unit-cell coordinates according to the kagome geometry. 
In the junction with translation symmetry along the $x$-direction but broken in the $y$-direction, it is instructive to work in the hybrid real-momentum representation. To this end, we perform a partial Fourier transform and recast the inversion operation as
\begin{subequations}
\begin{align}
   c_{k_x,\ell,1,\sigma} \;&\rightarrow c_{-k_x,-\ell+1,1,\sigma},\;    \\  
   c_{k_x,\ell,2,\sigma} \;&\rightarrow e^{ik_x}c_{-k_x,-\ell+1,2, \sigma},\; \label{eq:invesion-b} \\  
   c_{k_x,\ell,3,\sigma} \;&\rightarrow c_{-k_x,-\ell,3,\sigma},\; 
\end{align}
\end{subequations}
where the $e^{i k_x}$ factor in Eq.~\eqref{eq:invesion-b} arises from the translation of sublattice ``2'' along the column index $j$ (the $x$ direction).
The action of inversion on creation operators follows analogously, with $i\rightarrow-i$.
For a bulk kagome system with homogeneous parameters and $\phi=0$, all the terms in the Hamiltonian respect inversion symmetry. 
In a finite Josephson junction, however, inversion symmetry will be  broken at the interfaces because the kagome unit cells at the boundaries do not map onto themselves. This geometric asymmetry already affects the local terms, such as $H_{\Delta}$ and $H_{\text{cAFM}}$, it becomes even more pronounced for the nonlocal SOC term. Explicitly, the inversion operation yields $\mathcal{I} H_{so} \mathcal{I}^{-1}= H_{so}+\Delta H_{\text{inter}}$,
where the additional interface terms
\begin{align}
    H_{\text{inter}} & = \sum_{k_x,\sigma,\sigma^\prime} \bm{s}_{\sigma\sigma^\prime} \cdot \big[ f_{21}(k_x)\; \bm{n}_{21}  \big( c_{k_x,\mathcal{L}+1,2,\sigma}^\dagger  \notag \\
    &\quad  \times  c_{k_x,\mathcal{L}+1,1,\sigma^\prime}^{\vphantom{\dagger}}  -  c_{k_x,-\mathcal{L},2,\sigma}^\dagger c_{k_x,-\mathcal{L},1,\sigma^\prime}^{\vphantom{\dagger}} \big) \notag \\
    &\quad + g_{31}(k_x)\; \bm{n}_{31}\; \big(c_{k_x,\mathcal{L},3,\sigma}^\dagger  c_{k_x,\mathcal{L}+1,1,\sigma^\prime}^{\vphantom{\dagger}} \notag \\ 
        &\quad -  c_{k_x,-\mathcal{L},3,\sigma}^\dagger  c_{k_x,-\mathcal{L}+1,1,\sigma^\prime}^{\vphantom{\dagger}} \big) \notag \\
    &\quad + g_{32}(k_x)\; \bm{n}_{32} \; \big(c_{k_x,\mathcal{L},3,\sigma}^\dagger c_{k_x,\mathcal{L}+1,2,\sigma^\prime}^{\vphantom{\dagger}} \notag \\
    &\quad - c_{k_x,-\mathcal{L},3,\sigma}^\dagger c_{k_x,-\mathcal{L}+1,2,\sigma^\prime}^{\vphantom{\dagger}}\big)  
    \big]  + \text{h.c.}
\end{align}
arise because inversion maps SOC bonds at one interface to bonds that do not exist at the opposite interface. Therefore, $H_{so}$ becomes intrinsically inversion-asymmetric when restricted to the finite junction. The junction thus explicitly lacks inversion symmetry, providing one of the essential prerequisites for a finite Josephson diode effect.

Time-reversal symmetry $\mathcal{T}$ is a local antiunitary symmetry that flips both spin and momentum~\cite{RevModPhys.88.035005,bernevig2013topological}. 
In the (spin $\otimes$ sublattice) basis $\Psi$, the time-reversal operator reads
\begin{align}
    \mathcal{T}=i s^y I_{3\times 3} \mathcal{K},
\end{align}
where the three-by-three unit matrix $I_{3\times 3}$ acts in sublattice space and $\mathcal{K}$ denotes the complex conjugation. For $\phi=0$, the terms $H_0$, $H_{so}$ and $H_\Delta$ are invariant under time reversal. However, the cAFM exchange term $H_{\text{cAFM}}$ filps sign, $\mathcal{T}H_{\text{cAFM}}\mathcal{T}^{-1}=-H_{\text{cAFM}}$. Thus, $\mathcal{T}$ is also explicitly broken in the cAFM order. This provides another necessary ingredient for realizing the Josephson diode effect.

Finally, we analyze mirror reflection about the kagome plane. This mirror symmetry flips in-plane spin components while preserving out-of-plane component,  $\mathcal{M}_z:(s^x,s^y,s^z) \rightarrow (-s^x,-s^y,s^z)$. 
Its matrix representation can be written as~\cite{GaoPRB2016,rodriguez2012symmetry}
\begin{align}
    \mathcal{M}_z=i s^z I_{3\times 3}.
\end{align}
As the magnetic moments in the cAFM lie in the plane,  $\mathcal{M}_z H_{\text{cAFM}} \mathcal{M}_z^{-1}=-H_{\text{cAFM}}$. 
Thus, the cAFM order breaks both $\mathcal{T}$ and $\mathcal{M}_z$ individually. However, it preserves their combined symmetry, as indicated by $(\mathcal{TM}_z) H_{\text{cAFM}} (\mathcal{TM}_z)^{-1}=H_{\text{cAFM}}$. 
To generate a Josephson diode, this combined $\mathcal{TM}_z$ symmetry must also be broken. The SOC provides precisely such a symmetry breaking. 
time-reversal $\mathcal{T}$ flips the spin and  $\mathcal{T}i\mathcal{T}^{-1}= -i$, leading to $\mathcal{T}H_{so}\mathcal{T}^{-1}=H_{so}$. In contrast, 
mirror reflection $\mathcal{M}_z$ flips the in-plane spin components and gives $\mathcal{M}_zH_{so}\mathcal{M}_z^{-1}=-H_{so}$. Therefore, the SOC intrinsically breaks the $\mathcal{TM}_z$ symmetry of the kagome system, $(\mathcal{TM}_z) H_{so} (\mathcal{TM}_z)^{-1}=-H_{so}$~\cite{Chen2014PRL}.

\subsection{Field-free Josephson diode effect} 

We now demonstrate the emergence of the Josephson diode effect in the kagome SC/cAFMs/SC junction. The Josephson supercurrent can be obtained directly from the free energy~\cite{Tinkham2004Book}. For numerical efficiency and to avoid finite-size effects associated with the superconducting leads, we alternatively compute it employing the surface Green's function approach~\cite{Sancho1985JPFMP,Asano01prb,SBZhang20PRB}. 

\begin{figure}[t]
\centering
\includegraphics[width=1\linewidth]{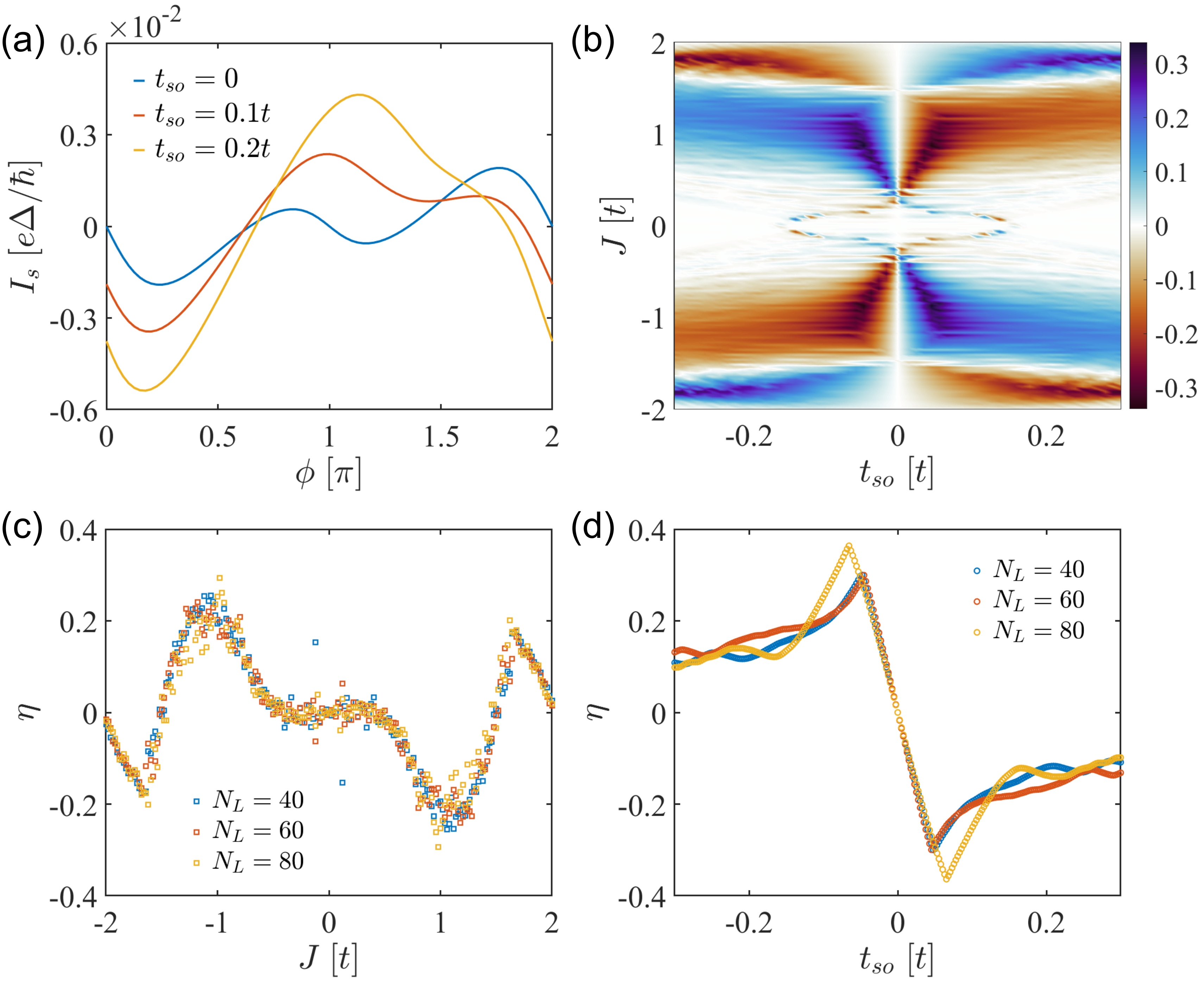}
\caption{Josephson diode effect in the SC/cAFM/SC junction with SOC. (a) Current-phase relations for cAFM strength $J=t$, junction length is $N_L=40$ and different SOC strengths $t_{so}=0$, $0.1t$ and $0.2t$. (b) Diode efficiency $\eta$ as a function of $J$ and $t_{so}$ for $N_L=40$. (c) $\eta$ as a function of $J$ for $t_{so}=0.1t$ and $N_L=40$, 60 and $80$. (d) $\eta$ as a function of $t_{so}$ for $J=t$,  $N_L=40$, $60$ and $80$. Other parameters are $\mu_{\text{AFM}}=0.2t$, $\mu_S=0.2t$, $\Delta=0.02t$ and temperature $k_BT = 0.02\Delta$. 
}
\label{fig:show_diode_effect_SOC}
\end{figure}

Figure~\ref{fig:show_diode_effect_SOC}(a) displays representative current–phase relations for a fixed cAFM strength ($J=t$) and several SOC strengths ($t_{so}=0$, $0.1t$, and $0.2t$). For illustration, we take the junction length $N_L=40$ (in units of row spacing $\sqrt{3}a$), set the chemical potentials of the cAFM and the SC to $\mu_{\text{cAFM}}=\mu_S=0.2t$, the superconducting pairing potential to $\Delta = 0.02t$, and the temperature to $k_BT = 0.02\Delta$. In the absence of SOC ($t_{so}=0$), the current-phase relation is anti-symmetric under $\phi\rightarrow -\phi$, indicating identical forward and backward critical currents and thus no diode effect. 
Strikingly, once a finite SOC ($t_{so}\neq 0$) is introduced, this antisymmetry is lifted and the forward and backward critical currents, $I_c^\rightarrow$ and $I_c^\leftarrow$, become unequal. The resulting diode efficiency, defined as $\eta\equiv (I_c^\rightarrow-I_c^\leftarrow)/(I_c^\rightarrow+I_c^\leftarrow)$, is therefore finite only when SOC is present. This demonstrates that the coexistence of SOC and cAFM order is capable of generating a Josephson diode effect in the kagome junction. 

To study the connection between the diode effect and the interplay between cAFM and SOC more carefully, we calculate the diode efficiency $\eta$ as a function of the cAFM strength $J$ and the SOC strength $t_{so}$, the results of which are presented in Fig.~\ref{fig:show_diode_effect_SOC}(b). Srikingly, a pronounced efficiency exceeding $30\%$ can be achieved. Furthermore, the efficiency is an odd function of both parameters, $\eta(-J)=-\eta(J)$ and $\eta(-t_{so})=-\eta(t_{so})$, and it vanishes whenever $J=0$ or $t_{so}=0$. These features show that the requisite symmetry breaking is provided cooperatively by the cAFM order and SOC, indicating that both ingredients are essential. They also suggest that the direction of the nonreciprocal transport can be controlled by reversing the sign of either $J$ or $t_{so}$.  

Figures~\ref{fig:show_diode_effect_SOC}(c) and \ref{fig:show_diode_effect_SOC}(d) further examine how the diode effect depends on the junction length $N_L$.
Specifically, Fig.~\ref{fig:show_diode_effect_SOC}(c) shows $\eta$ as a function of $J$ for a fixed $t_{so}=0.1t$ and increasing junction lengths, $N_L=40$, $60$, and $80$, while Fig.~\ref{fig:show_diode_effect_SOC}(d) shows $\eta$ as a function of $t_{so}$ for a fixed $J=t$ and the same set of junction lengths. Remarkably, although $\eta$ varies with both $J$ and $t_{so}$, it remains largely insensitive to $N_L$, demonstrating that the diode effect is robust and not a finite-size artifact. 
In addition, $\eta$ shows a nearly linear dependence on $t_{so}$ in the weak-SOC regime and then decreases slowly as $t_{so}$ becomes larger. Altogether, these results show that the interplay among the kagome lattice geometry, cAFM order, and SOC naturally produces a sizable Josephson diode effect even in the absence of external magnetic fields. We note that a similar diode effect driven by interfacial asymmetry has been recently discussed in a Josephson junction based on WTe$_2$~\cite{guo2025edge}. 



\subsection{Tunable $\phi_0$ junction state}

\begin{figure}[t]
\centering
\includegraphics[width=1\linewidth]{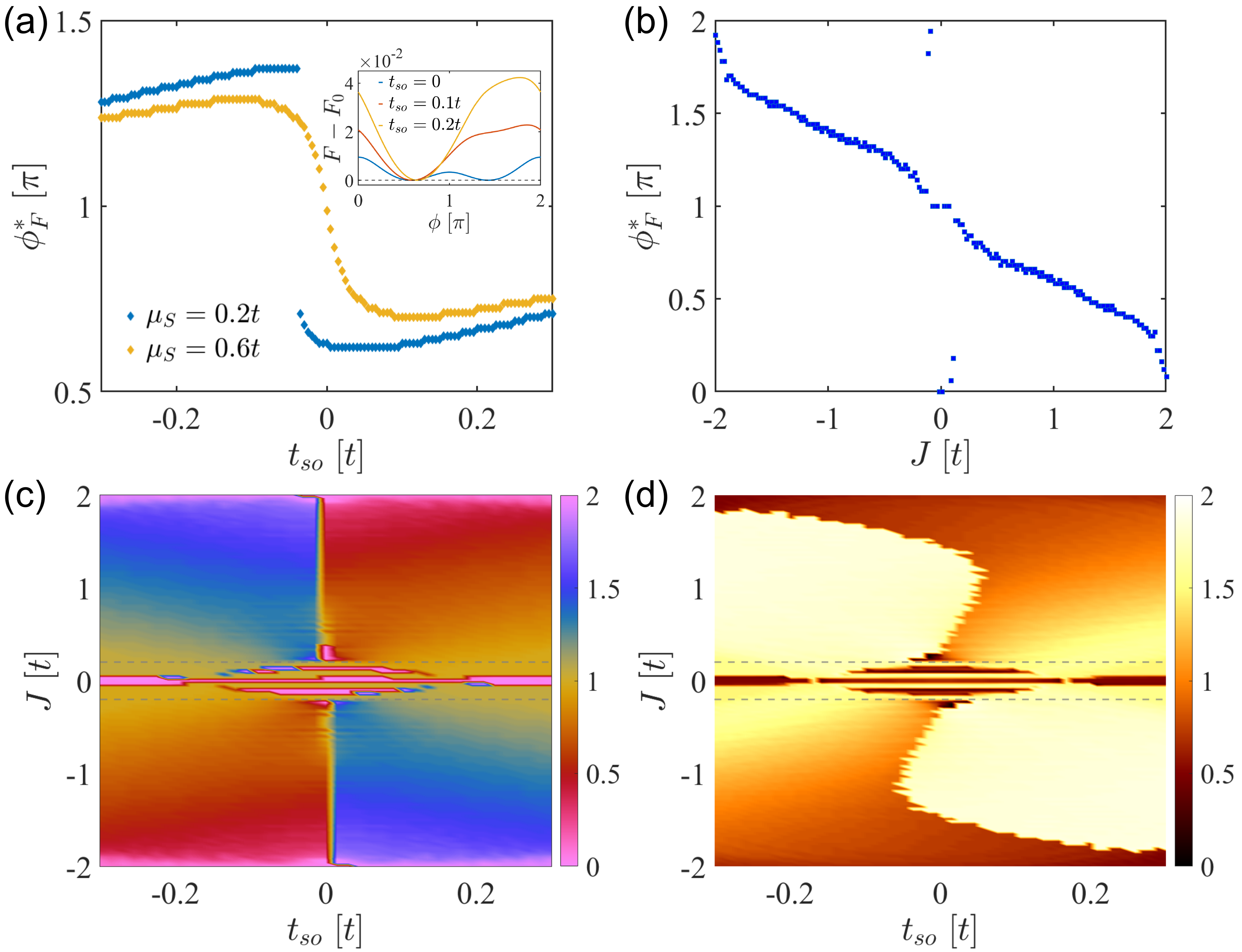}
\caption{$\phi_0$-junction state in the SC/cAFM/SC junction with SOC. (a) Phase position $\phi_F^*$ of the lowest free energy $F_0\equiv\min_\phi[{F(\phi)}]$ as a function of SOC strength $t_{so}$ for $J=t$, $\mu_S=0.2t$ (blue) and $\mu_S=0.6t$ (orange). Inset: free energy $F$ measured relative to $F_0$ as a function of the superconducting phase difference $\phi$ for $J=t$ and $t_{so}=0$, $0.1t$, $0.2t$. (b) $\phi_F^*$ of $F_0$ as a function of $J$ for $t_{so}=0.1t$ and $\mu_S=0.2t$. Phase diagram of (c) the phase position $\phi_F^*$ and (d) the phase position $\phi_I^{*}$ as functions of $J$ and  $t_{so}$ for $\mu_S=0.2t$. Other parameters are $\mu_{\text{AFM}}=0.2t$, $\Delta=0.02t$, $N_L=40$, and $k_BT = 0.02\Delta$. 
}
\label{fig:phi_junction_SOC_scenario}
\end{figure}

Beyond the diode effect, the kagome SC/cAFM/SC junction also supports a tunable $\phi_0$ state. In the presence of SOC, the phase position $\phi_F^*$ of the lowest free energy of the junction, $F_0\equiv\min_\phi[{F(\phi)}]$, is no longer restricted to the conventional values $\phi = 0$ and $\pi$. Instead, it can vary continuously over a broad range of phases under the combined influence of the cAFM order and SOC. This behavior is apparent from the evolution of the free-energy landscape as a function of the superconducting phase difference $\phi$. As illustrated in the inset of Fig.~\ref{fig:phi_junction_SOC_scenario}(a), the free energy $F(\phi)$ remains symmetric in the absence of SOC. Introducing a finite $t_{so}$ breaks this symmetry, causing the energy minimum $F_0$ to shift away from $\phi = 0$ and $\pi$ and thereby generating a $\phi_0$-junction state. Note that due to high interface transparency, the free energy may exhibit degenerate minima, causing abrupt changes in $\phi_F^*$ as $t_{so}$ or $J$ is varied. Reducing the transparency (e.g., by introducing a large Fermi-level mismath) suppresses higher-harmonic processes and lifts the free-energy degeneracy. 

Figure~\ref{fig:phi_junction_SOC_scenario}(a) displays the phase position $\phi_F^*$ as a function of SOC strength $t_{so}$ at a fixed cAFM strength ($J = t$) for two representative chemical potentials in the superconductor, $\mu_S = 0.2t$ (blue) and $\mu_S = 0.6t$ (orange). As expected, a finite anomalous phase $\phi_F^*$ deviating from $0$ and $\pi$ generally develops. For high interface transparency (e.g., with small Fermi-level mismatch $\mu_{\text{AFM}}=\mu_S=0.2t$), $\phi_F^*$ exhibits a sharp change at small $t_{so}$. In contrast,  for low interface transparency case (e.g., with large Fermi level mismatch  $\mu_{\text{AFM}}=0.2t$ and  $\mu_S=0.6t$), $\phi_F^*$ evolves smoothly with $t_{so}$. Figure~\ref{fig:phi_junction_SOC_scenario}(b) shows the dependence of $\phi_F^*$ on the cAFM strength $J$ for a given $t_{so}$. As $J$ increases, $\phi_F^*$ can be tuned across the full phase interval $[0,2\pi]$, further demonstrating the extensive tunability of the $\phi_0$-junction state. 

Figure~\ref{fig:phi_junction_SOC_scenario}(c) shows The phase diagram of the phase position $\phi_F^*$ as a function of the cAFM strength $J$ and SOC strength $t_{so}$. It demonstrates that $\phi_F^*$ is independently tunable by either parameter and exhibits an overall antisymmetric dependence on their signs, i.e., $\phi_F^*(-J,-t_{so})=\phi_F^*(J,t_{so})$. This establishes a broad and flexible control of the anomalous phase through magnetic order and spin--orbit interactions. To further connect this tunability with transport behavior, we compute the phase diagram of the phase position $\phi_I^*$ of the forward (positive) critical current. As shown in Fig.~\ref{fig:phi_junction_SOC_scenario}(d), $\phi_I^*$ shows a dependence on $J$ and $t_{so}$ that closely mirrors that of $\phi_F^*$, confirming that the $\phi_0$ shift is faithfully reflected in transport observables and can be extracted directly from the current--phase relation.
Note that both $\phi_F^*$ and $\phi_I^*$ exhibit noticeable oscillations with changing $J$ and $t_{so}$ when $J < |\mu_{\text{AFM}}|$, a regime where two Fermi surfaces are present at each valley. In contrast, when $J > |\mu_{\text{AFM}}|$, only a single Fermi surface remains per valley, and the phase positions evolve smoothly with $J$ and $t_{so}$.



\subsection{Diode effect in the SC/cAFM/cAFM$'$/SC junction induced by Zeeman fields}\label{sec:scenario:Mz}


To further demonstrate that breaking the combined $\mathcal{TM}_z$ symmetry, rather than SOC itself, is crucial for realizing the Josephson diode effect, we now propose and study an SC/cAFM/cAFM$'$/SC junction without SOC, as illustrated in Fig.~\ref{fig:setup}(b). 
From the previous setup, we have seen that, in the absence of SOC, inversion symmetry breaking induced solely by the junction interface is insufficient to produce a diode effect. To overcome this limitation, in the present second setup, we consider two cAFM regions in the junction with a relative angle $\theta^r$, which naturally breaks spatial inversion symmetry. Such a bi-cAFM configuration still preserves the $\mathcal{TM}_z$ symmetry and therefore cannot by itself generate a diode effect, as we will show below. This motivates us to further introduce a Zeeman exchange field into the junction. The total Hamiltonian of the junction is then given by
\begin{align}
 H_{{\text{SNNS}}}= H_0+H_\Delta+{\tilde{H}}_{\text{cAFM}}+H_{B}.
\end{align}
Here, the pairing term $H_\Delta$ is given by Eq.~\eqref{eq:pairing} but with a spatially depdent order paramter $\Delta_\ell=\Theta(\ell+N_{L_1})\Delta_0 e^{-i\phi/2} + \Theta(\ell-N_{L_2})\Delta_0 e^{i\phi/2}$. $N_{L_1}$ and $N_{L_2}$ denote the lengths of the two cAFM regions. Accordingly, the cAFM term ${\tilde{H}}_{\text{cAFM}}$ is given by Eq.~\eqref{eq:H_cAFM} but with 
$\bm{m}_{\ell,\nu}=\tilde{\bm{m}}_{1,\nu}\Theta(\ell+N_{L_1}+1)\Theta(-\ell+1) +\tilde{\bm{m}}_{2,\nu}\Theta(\ell)\Theta(-\ell+N_{L_2}+1)$.  
$\tilde{\bm{m}}_{\alpha,\nu} = J (\cos{\tilde{\theta}_{\alpha,\nu}},\sin{\tilde{\theta}_{\alpha,\nu}},0)$ is the magnetic moment at sublattice $\nu$ with strength $J$ and direction $\tilde{\theta}_{\alpha,\nu} = 2(3-\nu)\pi/3+\theta_\alpha^*$ in the $\alpha$-th  ($\alpha=1,2$) cAFM regions. The relative angle between the chiral anti-ferromagnetic orders in the two cAFMs is defined as $\theta^r\equiv\theta^*_2-\theta^*_1$. Finally, the Zeeman term is written 
\begin{align}
H_{B} = \sum_{\ell,\nu} \Psi^\dagger_{\ell,\nu}\ \bm{B}_\ell \cdot \bm{s}\ \Psi_{\ell,\nu}, 
\end{align}
where $\bm{B}_\ell=\bm{B}\Theta(\ell+N_{L_1}+1)\Theta(-\ell+N_{L_2}+1)$ and $\bm{B}=(B_x,B_y,B_z)$ is the Zeeman exchange field. 
An out-of-plane Zeeman field, $\bm{B}=(0,0,B_z)$, breaks $\mathcal{T}$ while preserving $\mathcal{M}_z$, and therefore violates the combined symmetry $\mathcal{TM}_z$. In contrast, an in-plane Zeeman field, $\bm{B}=(B_x,B_y,0)$,
breaks $\mathcal{T}$ and $\mathcal{M}_z$ individually, but each operation flips the in-plane spin once. Hence, the in-plane Zeeman field preserves $\mathcal{TM}_z$. 

\begin{figure}[h]
\centering
\includegraphics[width=0.96\linewidth]{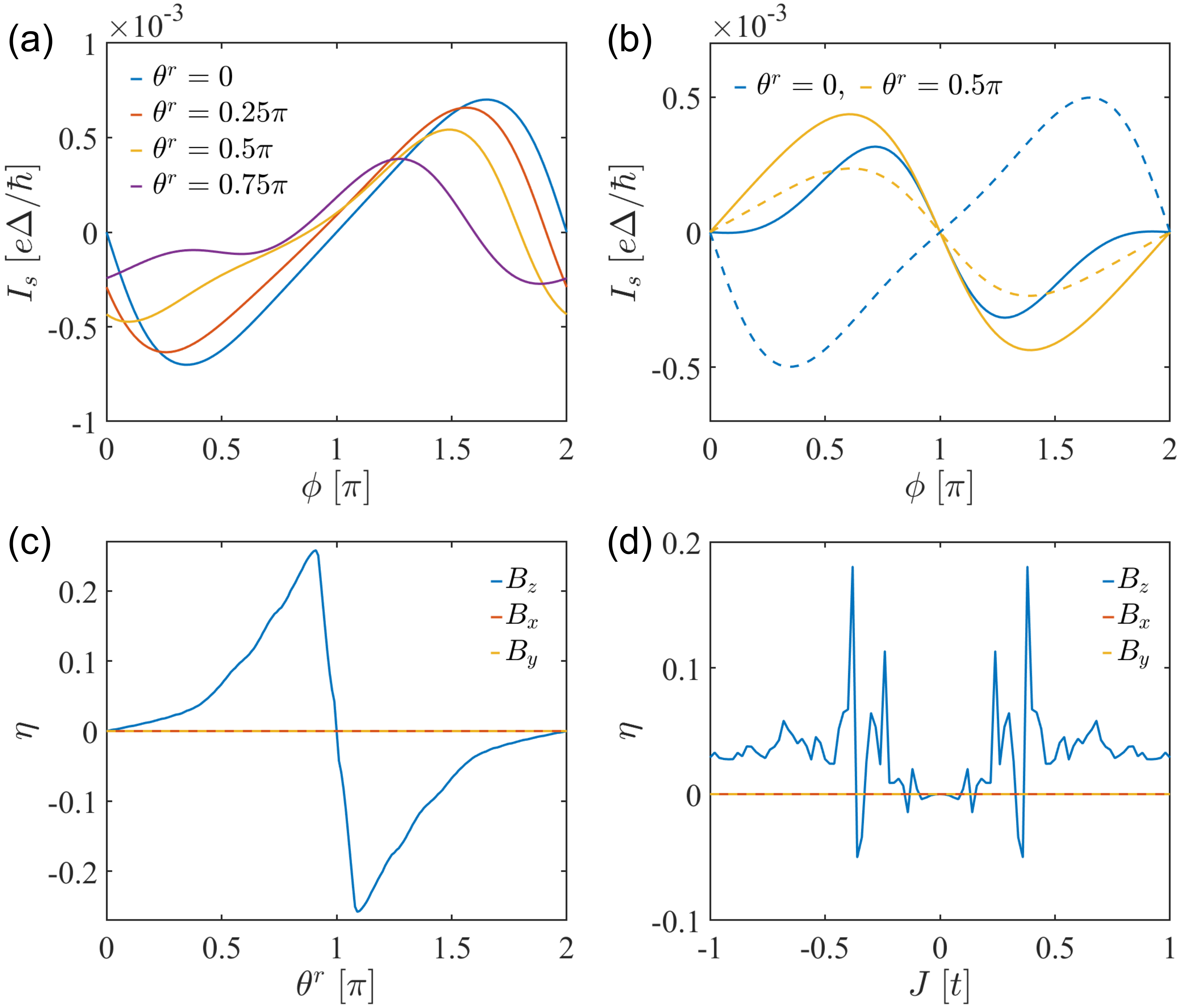}
\caption{Josephson diode effect in the SC/cAFM/cAFM$^\prime$/SC junction. (a) Current-phase relations for $J=0.4t$, under an out-of-plane Zeeman field $B_z=0.1t$ and for relative cAFM angles $\theta^r=0,\;0.25\pi,\;0.5\pi$ and $0.75\pi$. The corresponding diode efficiencies are $\eta=0$, $0.018$, $0.067$ and $0.173$, respectively. (b) Same as panel (a) but for an in-plane Zeeman field applied along the $x$-direction ($B_x=0.1t$, solid line) and the $y$-direction ($B_y=0.1t$, dashed line). No diode effect is observed for either in-plane field. (c) Diode efficiency $\eta$ as a function of the relative cAFM angle $\theta^r$ for $J=0.4t$ with Zeeman fields ${\bf B}= (0.1t,0,0)$, $(0,0.1t,0)$, and $(0,0,0.1t)$ in the $x$-, $y$-, and $z$-directions, respectively.  
(d) $\eta$ as a function of $J$ for $\theta^r = 0.5\pi$, with Zeeman fields ${\bf B}= (0.1t,0,0)$, $(0,0.1t,0)$, and $(0,0,0.1t)$ applied along the $x$-, $y$-, and $z$-directions, respectively. Other parameters are $\mu_{\text{AFM}}=0.2t$, $\mu_S=0.2t$, $\Delta=0.02t$, $N_{L_1}=N_{L_2}=20$, and $k_B T= 0.02\Delta$.}
\label{fig:DiodeEffect_ZeemanField}
\end{figure}

\begin{figure}[h]
\centering
\includegraphics[width=1\linewidth]{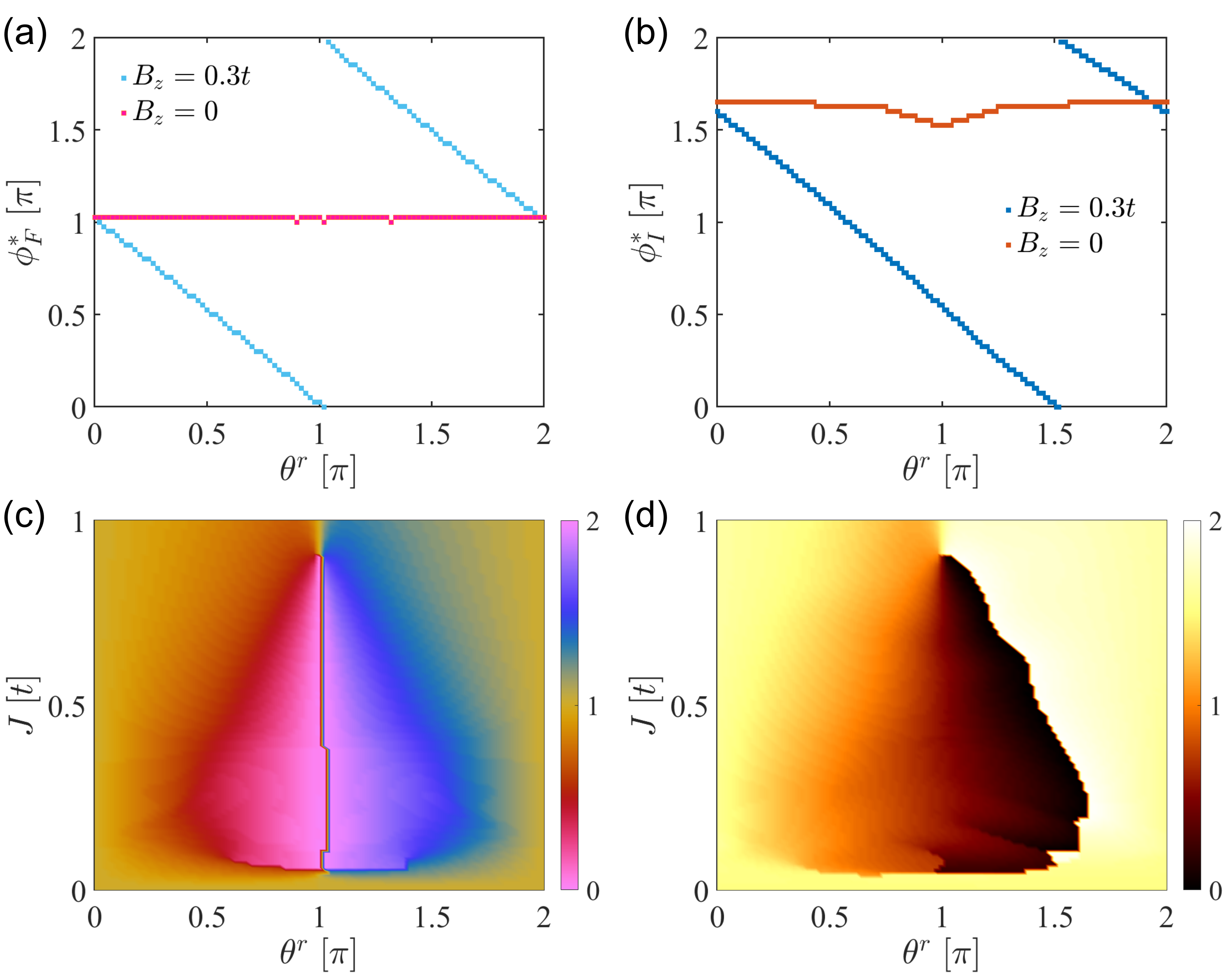}
\caption{$\phi_0$-junction state in the SC/cAFM/cAFM$^\prime$/SC junction. (a) Phase position $\phi_F^*$ of lowest free energy $F_0$ as a function of the relative angle $\theta^r$ between the two cAFM orders for $J=0.4t$ with ($B_z=0.3t$) and without ($B_z=0$) Zeeman field along $z$-direction. (b) Phase position $\phi_I^*$ of the forward critical current as a function of $\theta^r$ for $J=0.4t$ with ($B_z=0.3t$) and without ($B_z=0$) Zeeman field along $z$-direction. Phase diagram of (c) $\phi_F^*$ and (d) $\phi_I^*$ as functions of $J$ and $\theta^r$ for $B_z=0.3$. Other parameters are $\mu_{\text{AFM}}=0.2t$, $\mu_S=2t$, $\Delta=0.02t$, $N_{L_1}=N_{L_2}=20$, and $k_B T = 0.02\Delta$. 
}
\label{fig:fig6_phi_JunctionZeemanField}
\end{figure}


We discuss the Josephson current and how it can be tuned by the relative angle of the cAFM orders and the Zeeman field in the SC/cAFM/cAFM$^\prime$/SC junction. 
In the presence of an out-of-plane Zeeman field ${\bf B}=(0,0,0.1t)$, the current-phase relation is symmetric for $\theta^r=0$ or $\pi$, while becomes asymmetric when the cAFM orders acquire a relative angle $\theta^r$ (different from $\pi$, e.g., $0.25\pi$, $0.5\pi$, $0.75\pi$) [see Fig.~\ref{fig:DiodeEffect_ZeemanField}(a)]. Correspondingly, a pronounced diode effect develops (with the diode efficiencies $\eta=0.018$, 0.067, $0.173$, respectively). These results demonstrate that nonuniform cAFM order provides a natural source of inversion-symmetry breaking, while a Zeeman exchange field in the $z$-direction further breaks the $\mathcal{TM}_z$ combined symmetry, thereby enabling nonreciprocal supercurrent transport. In contrast, Zeeman fields applied in the $x$- or $y$- directions preserve $\mathcal{TM}_z$, leading to a symmetric current-phase relation and prohibiting the diode effect [see Fig.~\ref{fig:DiodeEffect_ZeemanField}(b)]. The $\theta^r$ dependence of the diode efficiency $\eta$ in Fig.~\ref{fig:DiodeEffect_ZeemanField}(c) highlights this symmetry constraint: $\eta$ is finite only for out-of-plane Zeeman fields and is odd in $\theta^r$. Moreover, it shows that the diode efficiency can be continuously tuned by changing the relative angle $\theta^r$ and can reach substantial values (exceeding 20\%). At $\theta^r=0.5\pi$, $\eta$ exhibits pronounced oscillations with cAFM strength $J$ for a fixed out-of-plane Zeeman field ($B_z=0.1t$) and is even in $J$, but remains negligible for in-plane Zeeman fields [see Fig.~\ref{fig:DiodeEffect_ZeemanField}(d)]. 
For a bilayer ferromagnetic counterpart, where the N\'eel vectors in the two ferromagnets have a relative angle, no diode effect is found. This comparison indicates that a non-collinear chiral antiferromagnetic order is essential for realizing the diode effect. 
Taken together, these findings establish that the Josephson diode effect in this system originates from the cooperative breaking of $\mathcal{I}$, $\mathcal{T}$, and $\mathcal{TM}_z$ symmetries by nonuniform cAFM order and an out-of-plane Zeeman field.

Finally, we investigate the tunability of the $\phi_0$-junction in the SC/cAFM/cAFM$^\prime$/SC structure. Figures \ref{fig:fig6_phi_JunctionZeemanField}(a,b) present the phase position $\phi_F^*$ of the lowest free energy $F_0$ and $\phi_I^*$ of the forward critical supercurrent as functions of the relative angle $\theta^r$ at fixed $J=0.4t$, both with and without an out-of-plane Zeeman field. In the absence of a Zeeman field ($B_z=0$, red markers), $\mathcal{TM}_z$ is preserved, $\phi_F^*$ remains pinned at $\phi_F^*=\pi$, and $\phi_I^*$ stays around $\phi_I^*=3\pi/2$, indicating a stable $\pi$ junction. By contrast, when an out-of-plane Zeeman field is applied ($B_z=0.3t$, blue markers), $\mathcal{TM}_z$ is broken, leading to a finite phase shift away from the normal values, $0 (\equiv2\pi)$ or $\pi$ for $\phi_F^*$ and $\pi/2, 3\pi/2$ for $\phi_I^*$. Moreover, both phase positions vary smoothly and nearly linearly  with $\theta^r$ across the full interval $[0, 2\pi]$, signaling the emergence of a tunable $\phi_0$-junction. To further characterize this tunability, figures~\ref{fig:fig6_phi_JunctionZeemanField}(c,d) present the phase diagrams of $\phi_F^*$ and $\phi_I^*$ as functions of $J$ and $\theta^r$. Both the phase positions $\phi_F^*$ and $\phi_I^*$ evolve continuously with $J$ and $\theta^r$, shifting away from the normal values. These results demonstrate that the $\phi_0$-junction is broadly tunable by both the cAFM strength $J$ and the relative angle $\theta^r$ under an out-of-plane Zeeman field.

\section{Conclusion and discussion}\label{sec:discussion}

We have investigated the Josephson diode effect and $\phi_0$-junction state in planar Josephson junctions incorporating $s$-wave superconductors and cAFMs on the kagome lattice. Our study demonstrates that both nonreciprocal Josephson transport and $\phi_0$-junction appear only when inversion $\mathcal{I}$, time-reversal $\mathcal{T}$, and combined mirror-time-reversal $\mathcal{TM}_z$ symmetries are simultaneously broken. Guided by this symmetry analysis, we propose two experimentally feasible Josephson setups that exhibit sizable diode efficiencies and tunable $\phi_0$ states. 

In the SC/cAFM/SC junction, the presence of SOC enables a field-free Josephson diode effect and a continuously tunable $\phi_0$ state. The diode response is tunable via the cAFM and SOC strengths, yet remains robust against variations in junction length, demonstrating its stability in extended systems. In the SC/cAFM/cAFM$^\prime$/SC setup, a relative angle between the cAFM orders in the cAFM regions, combined with an out-of-plane Zeeman field, gives rise to sizable diode responses and controllable $\phi_0$ junction states. These effects disappear under in-plane Zeeman fields or in the ferromagnetic counterparts, indicating the crucial role of $\mathcal{TM}_z$ symmetry breaking and highlighting the distinctive non-collinear character of cAFM order.
Taken together, our results establish a direct connection between $\mathcal{TM}_z$ symmetry breaking, cAFM textures, and nonreciprocal superconducting transport, positioning kagome cAFM systems as promising platforms for symmetry-engineered Josephson diodes and tunable $\phi_0$ junctions. 



\section{Methods}

The Josephson supercurrent \( I_s \) was determined by two complementary approaches: a thermodynamic formulation based on the system free energy and a microscopic treatment using the Green function formalism.  
Both schemes yield consistent results and are suitable for capturing the phase-dependent transport through superconducting junctions.

\subsubsection{Free-energy formulation}

In equilibrium, the supercurrent originates from the dependence of the total free energy \( F(\phi) \) on the superconducting phase difference \( \phi \) between the two superconductors. The current is given by 
\begin{equation}
I_s = \frac{2e}{\hbar} \frac{\partial F(\phi)}{\partial \phi},
\end{equation}
as established in the standard theory of Josephson junctions~\cite{Tinkham2004Book}.  
At finite temperature \( T \), the free energy is evaluated from the partition function \( Z = \sum_i e^{-E_i/(k_B T)} \) through \( F = -k_B T \ln Z \), where \( E_i \) denotes the many-body energy of the \( i \)-th state.  
In the zero-temperature limit, \( F \) reduces to the ground-state energy, obtained as the sum of all occupied (negative) single-particle eigenvalues of the system Hamiltonian.  
These eigenenergies are obtained via exact diagonalization of the Bogoliubov–de Gennes Hamiltonian for a given phase difference \( \phi \).

\subsubsection{Green-function formulation}

For systems with spatially dependent order parameters or magnetic textures, a Green-function treatment is numerically more efficient.  
Within the Matsubara formalism, the Josephson current can be written as~\cite{Asano01prb,SBZhang20PRB}
\begin{align}
I_s = & - \dfrac{ie}{2\hbar} k_BT \int\dfrac{dk_x}{2\pi} \sum_{\omega_n} \text{Tr} [\check{\tau}_3 \check{V}_+ \check{G}(\ell,\ell+1)  \notag \\ 
& \;\;\;\; \qquad \qquad \qquad \qquad -\check{\tau}_3 \check{V}_- \check{G}(\ell+1,\ell)],
\label{eq:current}
\end{align} 
where \( \omega_n = (2n + 1)\pi k_B T \) are the fermionic Matsubara frequencies.  
Here, \( \check{G}(\ell,\ell') \) is the Nambu Green function connecting two neighboring rows \( \ell \) and \( \ell^\prime \), \( \check{V}_\pm \) represent the inter-row hopping matrices (\( \check{V}_+ = \check{V}_-^\dagger \)), and $\check{\tau}_3$ is the third Pauli matrix in particle-hole space.  
The trace is taken over spin, particle–hole, and sublattice spaces.  
To simplify notation, the dependencies on \( k_x \) and \( \omega_n \) are omitted, i.e., \( \check{G}(\ell,\ell') \equiv \check{G}_{\omega_n}(\ell,\ell';k_x) \) and \( \check{V}_\pm \equiv \check{V}_\pm(k_x) \).  
The Green functions satisfy the recursion relations
\begin{subequations}
\begin{align}
\check{G}(\ell+1,\ell) &= \check{G}(\ell+1,\ell+1)\check{V}_+ \check{G}(\ell,\ell), \\
\check{G}(\ell,\ell+1) &= \check{G}(\ell,\ell)\check{V}_- \check{G}(\ell+1,\ell+1),
\end{align}
\label{eq:recGF}
\end{subequations}
which allow the full Green functions of the system to be constructed iteratively~\cite{Sancho1984JPFMP,Sancho1985JPFMP}.  

In the Nambu representation, the Green function at row \( \ell \) takes the block-matrix form
\begin{align}
\check{G}(\ell,\ell) =
\begin{pmatrix}
    \hat{G}_{e,\ell} & \hat{F}_{eh,\ell} \\
    \hat{F}_{he,\ell} & \hat{G}_{h,\ell}
\end{pmatrix},
\label{eq:Gfunction}
\end{align}
where \( \hat{G}_{e,\ell} \) and \( \hat{G}_{h,\ell} \) denote the normal Green functions for electrons and holes, respectively, while \( \hat{F}_{eh,\ell} \) and \( \hat{F}_{he,\ell} \) describe the anomalous (electron–hole) propagators that encode the superconducting correlations.  

The hopping matrices between adjacent rows can be written in the particle–hole basis as
\begin{align}
\check{V}_+ =
\begin{pmatrix}
    \hat{V}_e^+ & \hat{0} \\
    \hat{0} & \hat{V}_h^+
\end{pmatrix}, \qquad
\check{V}_- =
\begin{pmatrix}
    \hat{V}_e^- & \hat{0} \\
    \hat{0} & \hat{V}_h^-
\end{pmatrix},
\label{eq:HopMat_pm}
\end{align}
where \( \hat{V}_e^{\pm} \) and \( \hat{V}_h^{\pm} \) correspond to electron and hole hopping between neighboring rows.  

Substituting Eqs.~\eqref{eq:Gfunction} and \eqref{eq:HopMat_pm} into the general current expression [Eq.~\eqref{eq:current}] and the recursion relations [Eq.~\eqref{eq:recGF}], the Josephson current can be recast as
\begin{align}
I_s &=
 -\frac{ie}{2\pi\hbar} k_B T
 \int dk_x
 \sum_{\omega_n}
 \mathrm{Tr}\!
 \big[
   \hat{V}_e^+ \hat{F}_{eh,\ell} \hat{V}_h^- \hat{F}_{he,\ell+1} \notag \\
  & \qquad\qquad\qquad\qquad\qquad\;\; - \hat{V}_h^+ \hat{F}_{he,\ell} \hat{V}_e^- \hat{F}_{eh,\ell+1}
 \big].
\end{align}
This form highlights that the Josephson current is governed by the convolution of the anomalous Green functions \( \hat{F}_{eh} \) and \( \hat{F}_{he} \), which describe the pair correlations penetrating across the junction.  
In other words, the supercurrent results from the coherent transfer of Cooper pairs mediated by these anomalous propagators, with their spatial dependence reflecting the proximity and magnetic effects within the junction.

\begin{acknowledgments}
We thank James Jun He, Jian Li, and Alex Weststr\"om for helpful discussions. 
J.X.H. and S.B.Z. were supported by the start-up fund of HFNL, the Innovation Program for Quantum Science and Technology (Grant No. 2021ZD0302801), and the National Natural Science Foundation of China (Grant No. 12488101).
C.L. and L.H.H. were supported by the start-up fund of Zhejiang University and the Fundamental Research Funds for the Central Universities (Grant No. 226-2024-00068). 
C.L. was also supported by central fiscal special-purpose fund (Grant No.2021ZD0302500).
\end{acknowledgments}

{\emph{Note added:} When completing the manuscript, we became aware of a related preprint~\cite{frazier2025Spatially} which studies the Josephson diodes in non-collinear magnetic systems.}

\bibliography{diode_ref}

\end{document}